\documentclass[12pt]{article}

\usepackage{mathrsfs}

\begin{document}

    \title{Information and Particle Physics}

    \author{Wei Khim Ng\footnote{Email: phynwk@nus.edu.sg } \ and Rajesh R. Parwani\footnote{Email: parwani@nus.edu.sg}}

%\date{May 19 2007}

\maketitle

\begin{center}
%{\\}
{Department of Physics,\\}
{National University of Singapore,\\}
{Kent Ridge,\\}
{ Singapore.}

\end{center}

\begin{abstract}

Information measures for relativistic quantum spinors are constructed to satisfy various postulated properties such as normalisation invariance and positivity. Those measures are then used to motivate generalised Lagrangians meant to probe shorter distance physics within the maximum uncertainty framework. The modified evolution equations that follow are necessarily nonlinear and simultaneously violate Lorentz invariance, supporting previous heuristic arguments linking quantum nonlinearity with Lorentz violation.  The nonlinear equations also break discrete symmetries. We discuss the implications of our results for physics in the neutrino sector and cosmology.

\end{abstract}

\section{Introduction}
In statistical mechanics, the maximum entropy principle is a method to infer probability distributions that give the least biased description of the state of a system\cite{Jay,Buc,Buc1}. In a similar way one may understand the structure of the Schrodinger equation, in particular its linearity, by using maximum
entropy (uncertainty), or information-theoretic arguments \cite{Reg,Reg1}.

In the approach of
Ref.\cite{Reg,Reg1}, one starts with the action for classical ensemble dynamics, representing the coupled classical
Hamilton-Jacobi and continuity equations, and demands in addition that a certain measure of information, the
Fisher measure, is simultaneously minimised so as to maximise our uncertainty (minimise our bias) of the
microscopic dynamics. That procedure results in the usual Schrodinger equation after a change of variables
combines the two real coupled nonlinear equations into one linear complex equation for the wavefunction.

The use of the Fisher measure in Ref.\cite{Reg,Reg1} needs motivation: It can be constructed axiomatically \cite{Par1,Par1a} as the simplest measure satisfying constraints suitable to the context, just as the Gibbs-Shannon entropy measure is the simplest expression satisfying the requirements for statistical mechanics \cite{Jay,Sha}. Relaxing the constraints gives generalised measures\cite{Par2}, such as the Kullback-Liebler entropy\cite{kul}, which then lead within the maximum uncertainty approach to nonlinear Schrodinger equations whose  properties have been further investigated  \cite{Par3,Par3a,Par3b,Par3c}; an application to quantum cosmology \cite{Ngu,Ngu1} used the nonlinear equations to model expected new physics at the Planck scale: it was found that even a weak nonlinearity could replace the Big Bang singularity by a bounce.

This leads one to ask if a similar generalisation of the Dirac equation could be used to model, or probe for, new physics when the spin degree of freedom is relevant. In Ref.\cite{Ng1} nonlinear extensions of Dirac's equation were constructed axiomatically by requiring the extension to preserve various desirable properties of the original linear equation. Such equations, which generalise previous versions \cite{Doe,doe1,doe2,Bog,Bog1}, have since been used to study potential corrections to the standard neutrino oscillation probabilities \cite{Ng2} and may be relevant also in the condensed matter context \cite{Car}. In addition, the non-relativistic limit of the nonlinear Dirac equations \cite{Ng3} provides a natural regularisation of singularities found in previous studies of nonlinear Schrodinger equations \cite{Wei,Wei1}.

However, one drawback of the approach in Ref.\cite{Ng1} is that the constraints were not sufficiently restrictive, resulting in large classes of nonlinearities. This is because Ref.\cite{Ng1} did not use information-theoretic arguments as in Ref.\cite{Par2}:
While in the non-relativistic case one may start from a well-motivated classical system and obtain the linear or nonlinear information-theoretic extensions, in the relativistic case the corresponding classical starting point is unknown or ambiguous.

In this paper we adopt a strategy different from Ref.\cite{Ng1} to obtain generalisations of linear quantum evolution equations for spinors. As mentioned at the end of Ref.\cite{Par2}, once the usual linear quantum equations are available, one may consider using them as the new starting points for applying (again) an information-theoretic extension. That is precsiely what we do here for the relativistic equations, taking the spinor wavefunction $\psi(x,t)$ and its adjoint, rather  than the probability density, as the fundamental building blocks in the construction of information measures.

 Thus we wish to construct an action of the form
\begin{equation}\label{dirac1}
S = \int d^4 x (\mathscr{L}_0 + \mathscr{F} )
\end{equation}
where $\mathscr{L}_0=\bar \psi\left(i\gamma^\mu\partial_\mu-m\right)\psi$ is the usual starting Lagrangian leading to the linear Dirac equation while $I = \int d^4 x  \mathscr{F} $ is an information measure that is meant to be simultaneously minimised when deriving the equations of motion (the Lagrange multiplier method is used, the multiplier being implicit in $\mathscr{F} $). In this way one obtains generalised Dirac equations which we interpret as encoding potential new physics at higher energies. The positivity constraint on the information measure, to be discussed below, turns out to be very restrictive. {\it We emphasize that unlike \cite{Ng1}, here we do not start by demanding nonlinearity, but rather find it as one of the unavoidable consequences of an information-theoretic generalisation.}

As in Ref.\cite{Ng1} we work in this paper at the \emph{quantum mechanical level} rather than with quantum field theory. Alhough it is possible that our nonlinearities might include some standard quantum field theory corrections, we will explain in Sect.(5.1) why the form of $\mathscr{F}$ we obtain suggests other, more fundamental, corrections.

In the next section we outline and explain the conditions to be imposed on $\mathscr{F}$ so that $I$ may justifiably be called an information measure. Then in Sect.(3) we show that for Dirac spinors the conditions can only be satisfied if Lorentz invariance is violated. In Sect.(4) we discuss the minimisation condition and in Sect.(5) we give some examples of the nonlinear, Lorentz violating, Dirac equations. We also discuss the special cases of Weyl and Majorana spinors. In the concluding section we interpret our results and suggest how they might be used to probe for new physics at higher energies, involving also broken discrete spacetime symmetries.

Our notation and conventions for spinors are the standard ones used for example in Refs.\cite{Ng1,Pes}.

\section{Conditions}

We are interested in information measures, $I = \int d^4x \mathscr{F} $, constructed from the four component Dirac spinor $\psi$ and the adjoint $\bar \psi = \psi^{\dagger} \gamma^{0}$. We assume that $\psi, \bar\psi$ contract in the natural way in $\mathscr{F} $ to form scalars, for example $s_i=\bar \psi A_i \psi$, where $A_i$ is a matrix in spinor space which might contain derivatives and also depend on the wavefunction and its adjoint (contracted again in a similar way). The information measure should satisfy the following conditions:

\begin{itemize}

\item{[C1]} Homogeneity: The information measure should be homogeneous, that is invariant under the scaling $\mathscr{F}(\lambda\psi, \lambda \bar \psi )= \lambda^2 \mathscr{F}(\psi, \bar \psi)$, so that the modified evolution equation retains this property of the linear equation, allowing the wavefunction to be freely normalised: In this sense, our deformation is minimal. (An alternative motivation for this condition \cite{Ng1} is that for multiparticle states one desires the dimension of $\mathscr{F}(\psi)$ to be independent of the number of particles and hence the new coupling parameter (Lagrange multiplier) to be universal.)

\item{[C2]} Uncertainty: The information measure should decrease as $\psi(x,t)$ approaches a uniform value as then our uncertainty about the location of the quantum particle would be at a maximum \footnote{It is easy to check that $\psi^{\dagger} \psi$ is still the conserved probability density in our generalised equations \cite{Ng1}.}. We assume that $\mathscr{F}$ contains derivatives of $\psi$ that enforce this condition. Since $\mathscr{L}_0$ already contains derivatives, this appears to be a natural and simple solution.

\item{[C3]} Locality: All dependence of the wavefunction\footnote{Here and elsewhere obvious reference  to the adjoint is implied.} in $\mathscr{F}$ is at the same spacetime point and only a finite number of derivatives of the wavefunction occur. We assume therefore that $\mathscr{F} = {N \over D}$, where $N$ is a polynomial of the wavefunction containing a finite number of derivatives. The denominator $D$ is also a polynomial required to satisfy condition [C1].

\item{[C4]} Positivity: The information measure, which is an inverse uncertainty measure, should be non-negative for {\it generic} $\psi$. Thus $\mathscr{F} $ should be real and non-negative\footnote{A reminder for later: $\bar \psi \psi=\psi^{\dagger} \gamma^{0} \psi$ is a Lorentz scalar but it is not  positive definite.}.

\item{[C5]} Minimisation: The information measure should take a minimum value when one extremises the total action to obtain the equations of motion. This is required by the maximum uncertainty principle.

\end{itemize}

Conditions [C1] and [C3] are identical to those satisfied by the base Lagrangian $\mathscr{L}_0$, while Conditions [C2], [C4] and [C5] are required for an appropriate definition of an information measure and for its use within the maximum uncertainty framework. It is interesting to note that condition [C4] simultaneously guarantees that the extended equations remain Hermitian. As for separability for multiparticle states \cite{Ng1}, this is easily maintained by a class of $\mathscr{F}$ we consider in Sect.(4).

\section{Construction}

We start with the form suggested by condition [C3]. Then condition [C1] implies $\mathscr{F} = {N_{n+1} \over D_{n}}$, where $N$ and $D$ are polynomials constructed from the wavefunction and its adjoint. The positive integer subscripts $n$ and $n+1$ indicate the number of pairs of $\psi, \bar \psi$ that occur in each term of the corresponding polynomial; in order to satisfy [C2] and [C4], $n$ cannot be zero and so the information measure necessarily leads to nonlinear equations of motion.

Now, since there are derivatives in $N_{n+1}$ (and not in the denominator),  condition [C4]  implies that the numerator must be positive by itself, and so must be a sum of squares of real numbers, $N_{n+1} \sim \sum_{i} X_{p,i}^2$. Then, {\it in the absence of other fields}, the denominator must also be positive in a similar way, $D \sim \sum_{i} Y_{q,i}^2$; $p, q$ being integers which count the pairs of $\bar \psi, \psi$ in each term of $X,Y$ respectively. However the  squares in $N,D$ conflict with the homogeneity condition [C1]: together they imply the impossibility $2(2p-2q)=2$.

Thus one must implement positivity without making both the numerator and denominator of $\mathscr{F}$ a sum of squares. The first possibility is to make the denominator positive by using $\psi^{\dagger}$ instead of $\bar \psi$ when contracting it with $\psi$, and so giving rise to Lorentz violation.
The second possibility is to introduce such a Lorentz violating positivity in the numerator, and the third possibility is of course allow Lorentz violation in both the numerator and denominator.

Thus Lorentz violation is unavoidable if one uses Dirac spinors to maintain the positivity condition [C4] together with the homogeneity and locality conditions [C1], [C3]. This Lorentz violation can be written in terms of a background vector field, for example, $D_1= \bar \psi \gamma^{\mu} A_{\mu} \psi$ with $A_\mu=(A,0,0,0)$ in the frame where positivity is enforced. As in Ref.\cite{Col,Col1}, such covariant looking terms are not invariant under particle Lorentz transformations. On the other hand, under observer Lorentz transformations \cite{Col,Col1}, only those observers which are purely rotated with respect to the initial frame can interpret the generalised action in information-theoretic terms.

\section{Minimisation}
We have yet to investigate the fifth condition [C5]. Instead of attempting to construct the most general $\mathscr{F} = {N_{n+1} \over D_{n}}$ that satisfies [C5], we will simply display  a class of solutions.
Consider,

\begin{equation}\label{info-meas}
\mathscr{F}=\frac{\left(P(\bar \psi,\psi)\right)^{n+1}}{\left(Q(\bar \psi,\psi)\right)^{n}}, \label{class1}
\end{equation}
where $n \ge 1$ and $P,Q$ are each real numbers consisting of a (different) sum of bilinears in $\psi, \bar \psi$ such as
\begin{eqnarray}
P&=&a\bar \psi\psi+b_\mu\bar \psi\gamma^\mu\psi+c_{\mu\nu}\bar \psi\sigma^{\mu\nu}\psi+d_\mu\bar
\psi\gamma^\mu\gamma_5\psi\nonumber\\&&+e\bar \psi \gamma_5 \psi+\, \mbox{complex conjugate}.
\end{eqnarray}
where $a$ and $e$ are some constants or derivative operators while $b_\mu$, $c_{\mu\nu}$ and $d_\mu$ are some vector fields or derivatives.

Consider the variation  $\psi\rightarrow\psi+\epsilon\delta\psi$ of the action\footnote{As usual, in that variation $\bar \psi$ is treated as an independent variable and kept fixed.} about a soluton of the equations of motion. We write $P(\bar \psi,\psi)\rightarrow P(\bar \psi,\psi)+\epsilon P(\bar \psi,\delta\psi)\equiv P+\epsilon P'$ and $Q(\bar \psi,\psi)\rightarrow Q(\bar \psi,\psi)+\epsilon Q(\bar \psi,\delta\psi)\equiv Q+\epsilon Q'$. The real parameter $\epsilon$ keeps track of the order of infinitesimals, and the deviation $\delta \psi$ is chosen to maintain reality, that is, $P'$ and $Q'$ are real. The change in the total Lagrangian (\ref{dirac1}), to second order in $\epsilon$ is\footnote{We have labelled $\mathscr{L}(\bar \psi,\psi+\epsilon\delta\psi)$ and $\mathscr{L}(\bar \psi,\psi)$ by $\mathscr{L'}$ and $\mathscr{L}$ respectively.}

\begin{eqnarray}\label{min}
\Delta\mathscr{L}&=&\mathscr{L'}-\mathscr{L}=\frac{\epsilon^2n(n+1)P^{n-1}}{2Q^{n+2}}\left(PQ'-QP'\right)^2\nonumber\\&+&O(\epsilon^3) \, .
\end{eqnarray}
The terms of order $\epsilon$ have vanished  at the extremum which gives the nonlinear equations of motion. Since the $O(\epsilon^2)$ term (\ref{min}) is only due to the information measure (\ref{info-meas}), it is minimised at the extremum of the action, as required by condition [C5], if
\begin{eqnarray}
\mbox{for $n$ odd,} \,&& Q(\bar \psi,\psi)\,\, \mbox{is positive} \,  ,\label{odd} \\
\mbox{while for $n$ even,} \,&& P(\bar \psi,\psi)\,\, \mbox{is positive} \label{even} \, .
\end{eqnarray}

Notice that the restrictions (\ref{odd},\ref{even}) also make the information measure (\ref{info-meas}) positive and so are consistent with condition [C4]. Furthermore, by choosing $Q$ to be a  bilinear, separability for multiparticle states is easily achieved \cite{Ng1}.

Some attempted generalisations of (\ref{class1}) do not work: For example, for $n=1$ let the numerator of (\ref{class1}) be the product of two different bilinears $P_1,P_2$. Upon varying the action, we get
\begin{equation}
\Delta \mathscr{L}=\epsilon^2 \frac{(P_1'Q_1-P_1Q_1')(P_2'Q_1-P_2Q_1')}{Q_1^3}
\end{equation}
which need not be positive .

Finally, we remark that the positivity condition [C4] does not imply the minimisation [C5].  As a counter example, consider
\begin{equation}
R(\bar \psi,\psi)\frac{\left(P(\bar \psi,\psi)\right)^2}{\left(Q(\bar \psi,\psi)\right)^2}
\end{equation}
where $P,Q,R$ are different bilinears with $R$ positive definite. Upon the variation $\psi\rightarrow\psi+\epsilon\delta\psi$, and using notation similar to before,
\begin{equation}
\Delta\mathscr{L}= \epsilon^2 \frac{(QP'-PQ')\left[2PQR'+R(QP'-3PQ')\right]}{Q^4}
\end{equation}
which need not be positive in general.

\section{Explicit Examples}
In this section we present a few explicit examples of nonlinear Lagrangians within the class (\ref{info-meas}) that satisfy the conditions [C1]-[C5] in addition to being Hermitian and separable for multiparticle states.

\subsection{Nonlinear Dirac Lagrangian}

For $n$ odd, since we require $Q(\bar \psi,\psi)$ to be positive, the only bilinear that satisfies this condition is given by $A_\mu\bar \psi\gamma^\mu\psi$ where $A_\mu=(A,0,0,0)$, is a time-like constant positive background field. Thus Lorentz invariance is violated. Since the power of the bilinear $P(\bar \psi,\psi)^{n+1}$ is even, it is easy to check that all such information measures for Dirac spinors are odd under charge conjugation, CP transformation and an overall CPT transformation. This class therefore illustrates the general result: Breaking of CPT symmetry implying the breaking of Lorentz symmetry \cite{Gre}.

A specific example with $n=1$ is given by
\begin{equation}\label{eg1}
\mathscr{L}_1=\bar \psi\left(i\gamma^\mu\partial_\mu-m\right)\psi+\frac{\left(i\bar \psi\gamma^\nu\partial_\nu\psi-i(\partial_\nu\bar \psi)\gamma^\nu\psi\right)^2}{4A_\mu\bar \psi\gamma^\mu\psi}.
\end{equation}

Another example, again with $n=1$, is
\begin{equation}\label{eg2}
\mathscr{L}_2=\bar \psi\left(i\gamma^\mu\partial_\mu-m\right)\psi+\frac{\left(B_\nu\partial^\nu(\bar \psi\psi)\right)^2}{A_\mu\bar \psi\gamma^\mu\psi} \,
\end{equation}
where $B_\nu=(0,\mathbf{B})$ is a constant space-like background field. This example is interesting because taking its non-relativistic limit reduces it to the Fisher measure discussed in Refs.\cite{Reg,Reg1,Par1,Par1a}.

Standard Lorentz covariant quantum field theory generates nonlinear effective Lagragians which are covariant. However, as our quantum mechanical forms, such as (\ref{eg1}), violate Lorentz invariance, they are mostly modeling a different kind of new physics. If one wishes, by relaxing some of the conditions in Sect.(2), such as homogeneity,  one may construct Lorentz covariant information measures even for Dirac particles; the corresponding actions might then be effective descriptions of conventional physics.

\subsection{Weyl and Majorana particles}

Weyl spinors \cite{Pes} may be used to represent massless fermions. Since Weyl spinors, $\psi_w$ have only two components, we can repeat the arguments used in the Dirac case but with $\bar \psi$ replaced by $\psi_w^\dag$. Although $Q(\psi_w^\dag,\psi_w)=\psi_w^\dag\psi_w$ is positive, it is not Lorentz invariant (the two spinors are of the same handedness). Thus just as in the Dirac case, an example such as
\begin{equation}\label{eg3}
\mathscr{L}_3=i\psi_w^\dag\bar \sigma^\mu\partial_\mu\psi_w+\frac{\left(i\psi_w^\dag\bar \sigma^\mu\partial_\mu\psi_w-i(\partial_\mu\psi_w^\dag)\bar \sigma^\mu\psi_w\right)^2}{4\psi_w^\dag\psi_w} \, ,
\end{equation}
where $\bar \sigma^\mu=(I,-\sigma^i)$ with $I$ the identity matrix and $\sigma^i$ the conventional Pauli matrices, will break Lorentz invariance while attempting to satisfy the other conditions (As before, one may write $\psi_w^\dag \psi_w=A_\mu\psi^\dag_w\bar \sigma^\mu\psi_w$ with $A_\mu=(A,0)$).

Historically, the masslessness of the neutrino, and its minimal representation by the Weyl equation, provided a conceptually appealing understanding of parity violation. Although neutrino masses are currently the conventional explanation for neutrino oscillations, there is still no direct proof of neutrino masses. So one may ask if neutrino ``masses" might actually be purely energy dependent parameters \cite{Par2}, which vanish as $E \to 0$, so that neutrinos might essentially be Weyl fermions. It seems difficult to fit current data to the simplest possibility represented by (\ref{eg3}) but perhaps generalisations, involving a sum of terms or higher orders, might make this possible. Of course data from future experiments should reveal whether or not the neutrino mass is really an energy dependent parameter.

Neutrinos might be Majoranna particles \cite{Pes}, represented by massive spinors which are their own charge conjugate. However we find that this does not help in constructing information measures satisfying conditions [C1]-[C5] which preserve Lorentz invariance.

\section{Discussion}
The Standard Model of particle physics, though immensely succesful, still leaves a number of questions unanswered, such as the reason for the smalless of neutrino masses. As a consequence many ``beyond the standard model" approaches have been attempted in the literature, for example invoking extra dimensions or fields, to provide some answers. The different framework we have investigated here, the information-theoretic approach, has been successfully used in other domains \cite{Buc,Buc1}. Instead of assuming a specific dynamics for the hypothesized new microphysics, one uses existing variables (in this case the wavefunction) to parametrise our uncertainty of the potential new physics. 

Compared to our previous study \cite{Ng1}, the present information-theoretically motivated approach to generalising  quantum  evolution equations for spinors gives more restrictive structures with fewer assumptions. Firstly, we find that while the basic linear equations are first order in derivatives, the  generalisations are necessarily nonlinear and involve higher derivatives. These higher derivative equations will then imply extra degrees of freedom, corresponding to more massive modes in addition to the usual modes of the linear equation \cite{Ng1}; these modes should manifest themselves at higher energies.

Our information-theoretic framework further suggests that if there is any new physics at higher energies corresponding to a quantum mechanical nonlinearity, then it is  simultaneously Lorentz violating. We think, as in Ref.\cite{Par2}, that the converse is likely to be true: Lorentz violation implying quantum mechanical nonlinearity.

We re-iterate that, in this paper, we demanded neither nonlinearity nor Lorentz violation of our Dirac equation extensions: Those properties turned out to be {\it simultaneous consequences} of the conditions [C1]-[C5] imposed within the information-theoretic framework. As discussed in Ref.\cite{Par2}, previous studies have treated those phenomena separately, looking at either quantum nonlinearity or Lorentz violation.

While in Ref.\cite{Ng2} we focused on neutrino oscillations at high energies to probe for quantum nonlinearities, we note that our nonlinear equations might also be useful for precision low energy tests of quantum linearity and violations of Lorentz and CPT symmetries. Taking (\ref{eg1}) as an example, one can show that the information measure, in the plane-wave approximation \cite{Ng1}, leads to a modified dispersion relation with an effective mass $m_{eff}=m[1-m^2/(AE)]$. The nonlinear effects increase as the energy decreases. Of course, if one simply wanted a phenomenological model with such a modified dispersion relation, one could simply replace the usual mass in the linear Dirac equation by the energy-dependent mass, leading to a linear, but highly non-local, equation. Clearly one can distinguish such linear models from our nonlinear equations by looking for specific nonlinear effects either at the non-relativistic level \cite{Wei,Wei1,Ng3} or possibly in future neutrino oscillation experiments \cite{Ng2}.

Another possible application of our results is towards understanding the baryon asymmetry in the universe. Our equations suggest a source of CP and CPT violation associated with quantum nonlinearities and Lorentz violation. We estimate the relevant energy scale from (\ref{eg1}) as follows: Let $\epsilon$ denote the strength of the nonlinearity represented by $1/A$. Since $\epsilon$ has a dimension of length, we compare it with the natural Compton scale, $1/M$, of the linear theory and write $\epsilon \sim f/M$ where $f$ is the dimensionless size of Lorentz violation. Thus in the plane wave approximation \cite{Ng2} we approximate the size of the nonlinearity, $F = fE^2/Mc^2$.  Treating $F$ as a correction to the electron mass $m_e$ \cite{Par2}, and using $f \sim 10^{-31}$ we get $E \sim 10^{12} GeV$ if
$M \sim m_e$   and $E \sim 10^{15} GeV$ if $M$ is the electroweak scale which gives masses to the particles.
For some previous studies using CPT violation to motivate baryogenesis, see Ref.\cite{Ber,Ber1} and references therein.

Finally, the approach used here may also be used directly at the non-relativistic level to generate novel information-theoretically motivated nonlinear Schrodinger equations \cite{Ng4}, particularly those that are integrable \cite{Par4}.

\end{document}